\documentclass[conference,compsoc]{IEEEtran}
\IEEEoverridecommandlockouts

\usepackage{cite}
\usepackage{amsmath,amssymb,amsfonts}
\usepackage{algorithmic}
\usepackage{graphicx}
\usepackage{textcomp}
\usepackage{xcolor}
\usepackage{float}
\usepackage{caption}
\usepackage{setspace}

\usepackage[hyphens]{url}        
\usepackage{hyperref}             
\hypersetup{
    colorlinks=false,     
    hidelinks,            
    breaklinks=true       
}
\usepackage{etoolbox}
\appto\UrlBreaks{\do\-}           
\appto\UrlBreaks{\do\_}           
\appto\UrlBreaks{\do\/}           
\def\BibTeX{{\rm B\kern-.05em{\sc i\kern-.025em b}\kern-.08em
    T\kern-.1667em\lower.7ex\hbox{E}\kern-.125emX}}
\begin{document}

\title{Goal-Driven Risk Assessment for LLM-Powered Systems: A Healthcare Case Study\\
}

\author{\IEEEauthorblockN{ Neha Nagaraja}
\IEEEauthorblockA{\textit{School of Informatics, Computing, and Cyber Systems} \\
\textit{ Northern Arizona University}\\
Flagstaff, USA \\
nn454@nau.edu}
\and
\IEEEauthorblockN{Hayretdin Bahsi\textsuperscript{1,2}}
\IEEEauthorblockA{\textsuperscript{1}\textit{School of Informatics, Computing, and Cyber Systems} \\
\textit{Northern Arizona University}, Flagstaff, USA \\
\textsuperscript{2}\textit{Department of Software Science} \\
\textit{Tallinn University of Technology}, Tallinn, Estonia \\
hayretdin.bahsi@nau.edu}
}

\maketitle

\begin{abstract}
\setstretch{1.1}
While incorporating LLMs into systems offers significant benefits in critical application areas such as healthcare, new security challenges emerge due to the potential cyber kill chain cycles that combine adversarial model, prompt injection and conventional cyber attacks. Threat modeling methods enable the system designers to identify potential cyber threats and the relevant mitigations during the early stages of development. Although the cyber security community has extensive experience in applying these methods to software-based systems, the elicited threats are usually abstract and vague, limiting their effectiveness for conducting proper likelihood and impact assessments for risk prioritization, especially in complex systems with novel attacks surfaces, such as those involving LLMs.  In this study, we propose a structured, goal driven risk assessment approach that contextualizes the threats with detailed attack vectors, preconditions, and attack paths through the use of attack trees. We demonstrate the proposed approach on a case study with an LLM agent-based healthcare system. This study harmonizes the state-of-the-art attacks to LLMs with conventional ones and presents possible attack paths applicable to similar systems. By providing a structured risk assessment, this study makes a significant contribution to the literature and advances the secure-by-design practices in LLM-based systems.
\end{abstract}


\section{Introduction}
Large Language Models (LLMs) are increasingly being adopted across diverse domains to perform a wide range of tasks, including information retrieval, text generation, summarization, question answering, and decision support~\cite{openai2024gpt4}. In healthcare, LLMs to support clinical decision-making, facilitate patient communication, and automate the summarization of medical records~\cite{denecke2024}. As AI becomes foundational to critical infrastructure and sensitive sectors like healthcare, securing them against adversarial threats is imperative. The 2025 America’s AI Action Plan~\cite{whitehouse2024ai} underscores this urgency, calling for secure-by-design AI and structured approaches to risk assessment and vulnerability identification.

Despite gains in efficiency and personalized care, integrating LLMs into safety-critical healthcare workflows introduces severe security challenges. Unlike traditional software, LLMs exhibit emergent behaviors, complex interactions, and are vulnerable to adversarial manipulation across inputs and internal states~\cite{zhui2024}. This exposes them to both conventional and AI-specific security threats. Adversaries may manipulate behavior via prompt injection~\cite{Rossi2024} or jailbreak~\cite{Chu2024}, extract training data through model inversion or membership inference, or leak EHR content via session memory. These threats not only compromise the integrity and availability of healthcare services but also raise serious concerns about patient privacy, regulatory compliance, and overall system trustworthiness. 

Recent efforts including the OWASP Top 10 for LLMs~\cite{OWASP2024} aim to identify and classify security threats in LLM-powered systems. However, the field still lacks the demonstration of structured risk assessment studies that connect identified threats to risks with real-world impact. In our prior work~\cite{ICISSP2025}, we identified and classified threats across key components of an LLM-based healthcare system using a STRIDE-based methodology, establishing a comprehensive foundation for understanding security posture and requirements. 

However, while prior work, including our own, has laid the groundwork for identifying threats in LLM-based systems, there remains a critical gap in how these threats are assessed in terms of risk. Existing studies~\cite{Pankajakshan2024,Tete2024} tend to stop at threat enumeration or qualitative assessments, without offering a structured framework to evaluate which threats are most severe, under what conditions they emerge, or how they can be systematically prioritized. Identified threats are usually abstract and vague, without being linked to potential attack scenarios and paths, which complicates the likelihood assessment of threats. It is difficult to perceive their impacts, as threats are individually elicited for each system component without being equipped with relevant information on how they induce the final impact. Thus, the lack of contextualization for the elicited individual threats does not lead to structured risk evaluation practices, reducing the practical utility of existing threat models, particularly in healthcare, where LLM errors can jeopardize patient safety, clinical accuracy, and regulatory compliance. 

To address the current gap in structured risk assessment for LLM-based systems, we present a goal-driven risk modeling framework, demonstrated in the context of healthcare.  Departing from the threats identified for each system component in the threat modeling phase, our method utilizes attack trees to contextualize individual threats to achieve three clinically grounded attacker goals: (G1) Intervening in Medical Procedures, (G2) Leakage of EHR Data, and (G3) Disruption of Access or Availability, in a healthcare system. Each attack tree demonstrates how a combination of several adversarial actions, conventional cyber attacks, LLM-specific attacks (e.g., prompt injection) or adversarial model attacks (e.g., model extraction), could achieve the given high-level goals through concrete attack scenarios and paths. Thus, our method extends beyond threat enumeration, providing an enriched context to better assess the likelihood and impact of threats for more rigorous risk scoring practices. This paper makes a significant contribution by presenting the first rigorous, structured risk assessment specifically tailored to LLM-based healthcare systems. 
\section{Related Work}

Although LLMs are rapidly adopted in critical domains like healthcare, structured methods to assess their security risks remain scarce. Prior studies have proposed threat taxonomies for LLMs, covering behaviors like prompt injection, data leakage, and hallucination, using STRIDE or stakeholder-centric views~\cite{Tete2024,Pankajakshan2024,hamid2024, clusmann2024}. While informative, these efforts focus on classification and rarely link threats to systemic impact. Traditional risk assessment frameworks have been applied in healthcare IT domains, including wireless networks and medical infrastructure~\cite{aijaz2023}. Hospital data show shared IT and market concentration cause cascading outages~\cite{yurcik2024}. Yet, these methods often rely on static, coarse grained views listing isolated issues without modeling how risks evolve or interact across components through attack scenarios. They omit attacker behavior, lack attack paths, and fail to capture how vulnerabilities propagate across workflows. As a result, they fall short in reflecting the full range of risks to clinical operations and patient safety~\cite{aijaz2023}.
Moreover, they overlook emergent LLM behaviors, context sensitive outputs, evolving memory states, and prompt susceptibility that complicate static threat enumeration and demand flexible, goal-driven risk models. Similarly, IoT- and embedded-device threat models~\cite{omotosho2019,vakhter2022,10538960} provide insight into cyber-physical risks but overlook AI-native threats and evolving software vulnerabilities.
Work on user trust and AI-driven decision-making~\cite{298114} focuses on patient–AI interactions and micro-level safety, but omits system-level risk modeling. Specifically, they fail to trace how threats propagate across broader infrastructure or interconnected components. 

Recent studies have moved beyond taxonomies and scoring systems by applying structured models like attack-defense trees to domains such as ATM systems~\cite{fraile2016} and ML pipelines~\cite{hoseini2024}. However, these frameworks remain largely unexplored for the unique risks posed by LLMs in safety-critical settings. Schiele et al.~\cite{schiele2025} demonstrated that attack-defense trees can support explainable risk reasoning, even for users with limited technical background. While their work focused on defense-oriented modeling, it highlights the broader potential of tree-based approaches for transparent security analysis. In this study, we focus on the attack modeling side and apply structured attack trees to assess and prioritize risks in LLM-based healthcare systems. This approach lays the groundwork for future extensions toward defense planning and risk remediation. 

In contrast to prior studies, we propose a goal-driven risk assessment framework grounded in system-level modeling and clinical impact. By combining attack tree with Likelihood × Impact scoring, our approach bridges the gap between qualitative threat taxonomies and practical, security-by-design reasoning for LLM-based healthcare systems.

\section{Methodology}

\begin{figure*}[ht]
    \centering
    \includegraphics[width=0.95\textwidth]{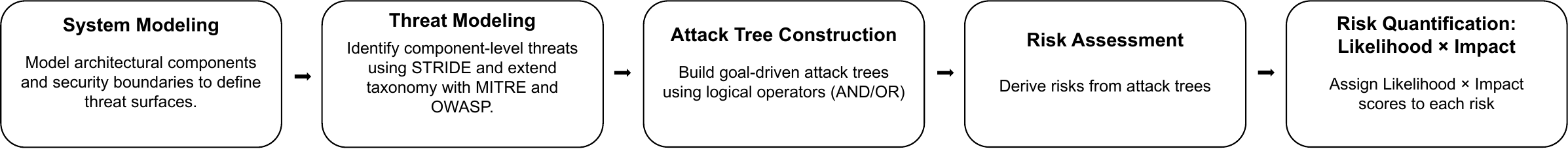}
    \caption{Overview of the proposed risk assessment methodology}
    \label{fig:methodology}
\end{figure*}

To structure our risk assessment, we adopt a layered methodology integrating system modeling, threat elicitation, attack tree construction, and risk quantification. Figure~\ref{fig:methodology} illustrates this process. We begin by modeling the architecture and trust boundaries of the LLM-based healthcare system to define its threat surface. Next, we apply established frameworks; STRIDE, MITRE ATLAS, and OWASP Top 10 for LLMs to identify threats at the component level. These threats are then mapped into goal-driven attack trees, which outline the logical steps an adversary may take to achieve specific objectives. From these trees, we derive goal-specific risks, modeling their evolution from initial conditions (preconditions) to final impact. Finally, we assess each risk using a structured Likelihood × Impact framework, enabling effective prioritization. The following subsections detail each stage of this process.

\subsection{System Modeling}
To support structured risk assessment, we model a representative LLM-powered healthcare system that handles sensitive data, provides medical guidance, and integrates third-party services. Each component introduces distinct functions and potential vulnerabilities, forming the foundation for our attack tree–based risk evaluation. This architecture builds on our prior work~\cite{ICISSP2025}, where we applied STRIDE for threat modeling. Here, we revisit it with a risk-oriented lens to highlight the security relevance of each component.

The architecture consists of five core components: \textbf{Web Application}: Main interface for submitting natural language queries; forwards inputs securely and initiates session flow. \textbf{Healthcare Platform}: Stores EHRs and provides clinical data to the Web App and Orchestrator.
\textbf{Orchestrator}: Serves as the system’s coordination layer. It functions as an LLM agent, managing the end-to-end interaction across internal modules and external services. It consists of \textit{Task Executor} which parses and routes user queries, manages API calls, and initiates system actions.
\textit{Task Planner}, which generates high-level strategies and prepares structured prompts for the LLM.
\textit{Data Pipeline}, stores intermediate data for reasoning. \textbf{External Resources}: APIs for translation, clinical databases, and analytical tools supporting data enrichment and multi-language support. \textbf{Large Language Model (LLM)}: Core reasoning engine that generates context-aware responses from structured prompts. 

\textbf{Assumptions}: While prior threat modeling assumed inference-only LLMs, this risk assessment considers scenarios involving fine-tuning or adaptation, reflecting real-world healthcare practices and expanding the threat surface.

\subsection{Threat Modeling}

Our threat modeling builds on prior work~\cite{ICISSP2025}, where we applied the STRIDE-per-element methodology to identify threats across the system architecture. Each component including processes, data flows, data stores, and external entities was analyzed for potential threats relevant to LLM-powered healthcare environments.

In STRIDE modeling, trust boundaries mark zones with different privilege levels. We defined eight: User, Healthcare Platform, External Resources, Orchestrator, LLM Interaction, External Communication, Model Management, and Data Logging. Cross-boundary flows were treated as potential attack paths, while intra-boundary flows were assumed secure due to mutual trust.

We compiled a comprehensive attack taxonomy using the MITRE ATLAS framework, extending it with critical entries from the OWASP Top 10 for LLMs~\cite{OWASP2024} and traditional cybersecurity threats. To support risk structuring, we categorized the identified the threats into three types~\cite{ICISSP2025}: (1) Conventional cyber threats that target infrastructure (e.g., MitM, unauthorized access); (2) Adversarial ML threats that exploit or subvert underlying models; and (3) Conversational threats that manipulate LLM behavior via inputs or context (e.g., prompt injection). The threats identified in this prior analysis form the foundation for the attack tree construction and risk assessment presented in this study. By reusing these findings, we model how threats propagate across components and inform risk prioritization.

\subsection{Risk Assessment}

Building on our previous threat modeling, we now present a risk assessment methodology that prioritizes risks by impact and likelihood. Instead of evaluating individual threats,  we derive structured risk evaluations from the goal-specific attack trees developed in this study. These trees capture how adversarial actions propagate through the system under specific conditions, allowing us to analyze each risk’s severity and the systemic consequences it may trigger.

\textbf{Attack Tree (AT) Modeling}: To guide structured risk analysis, we constructed dedicated attack trees for each high-level security goal: \textit{G1: Intervening in Medical Procedures}, \textit{G2: Leakage of EHR Data}, and \textit{G3: Disruption of Access or Availability}. Each tree models the various ways an adversary can achieve the goal, representing the attacker’s objective as the root node and tracing multiple attack paths through intermediate conditions and subgoals. Internal nodes correspond to necessary preconditions or attack steps, organized using logical operators: AND nodes capture dependencies that must all be satisfied, while OR nodes represent alternative means to progress. Leaf nodes represent atomic attack steps or system-level actions. These trees are grounded in threats identified in our prior threat modeling work. This structure allows us to analyze how specific threats propagate across the system and which conditions elevate a threat’s feasibility.

To further support actionable risk estimation, we model how threats evolve over time. Preconditions to Final Impact: Risk is not static; it evolves as attackers progress through various system stages. To support actionable risk estimation, we model threats along a timeline of attacker goals, capturing their structured evolution—essential for overcoming the limitations of vague or isolated threat descriptions. By contextualizing each threat with its preconditions (what must be true for it to occur), execution phase (how the attack unfolds), and final impact (what consequence it produces), we transform abstract threat definitions into concrete, analyzable attack scenarios. \textit{1. Preconditions}: the set of system states, vulnerabilities, or assumptions that must hold for an attack to be feasible. For example, for indirect prompt injection, the presence of a maliciously embedded instruction in external resources is a critical enabler. \textit{2. Execution Phase}: the observable attack behavior e.g., submitting a malicious input, injecting a task, hijacking a session. \textit{3. Final Impact}: the ultimate result or system compromise(e.g., incorrect diagnosis, EHR data leakage, system downtime). This phased structure allows us to trace how individiual threats can lead to high-severity system compromises, enabling early threat interception. Furthermore, it clarifies inter-dependencies across threat types. For example, consider a threat like indirect prompt injection. In a flat description, this may be summarized as “an attacker injects a malicious prompt via an external knowledge base.” However, using our structured threat evolution model: The precondition is that the system uses external plugins or resources (e.g., a translator or knowledge base) without sanitization. The execution phase involves an attacker inserting a hidden instruction into a trusted external content source. The final impact is unauthorized intervention in medical procedures (e.g., LLM recommending a contraindicated drug). This decomposition reveals that indirect prompt injection shares a precondition (unsanitized third-party inputs) with risks like knowledge base poisoning, highlighting how different threats may stem from shared vulnerabilities. 

We do not treat each goal (e.g., G1) as a monolithic outcome. Instead, we break them down into risk instances (e.g., G1-R1 to G1-R4), each representing a distinct form of system compromise. These risks are derived from each goal’s attack tree by identifying sub-paths that result in high-impact outcomes (e.g., misdiagnosis, unauthorized procedures). Each risk aggregates specific attack vectors, mapped to Likelihood × Impact scores based on feasibility and severity. This allows for more precise risk prioritization and aligns with how real-world healthcare systems assess safety-critical and privacy-relevant threats.

By visualizing both attack progression and threat escalation, these ATs help identify: Critical preconditions that enable high-impact threats, risks that persist across multiple paths, attack paths that span multiple system components. This approach moves beyond traditional risk tables by capturing the dynamics and conditional logic of real-world multi-stage attacks. While threat elicitation was component-driven, the risk assessment is goal-driven, structured around the three security goals introduced above. For each goal, we identify all relevant attack paths.  For each risk instance (e.g., G1-R1), we consider all associated attack paths and identify the most feasible one based on attacker capabilities and system context. The Likelihood score is then assigned based on this dominant path, using our defined Likelihood framework Table~\ref{tab:likelihood_framework}. We assign risk scores (Likelihood × Impact) based on their severity and relevance to real-world deployment, We analyze the preconditions, dependencies, and likely impact zones. This dual mapping; component-to-threat and threat-to-goal ensures that no critical area of the system is overlooked. The goal-driven framing also aligns with real-world risk acceptance models in healthcare, where impact is often measured in terms of harm to patient care (G1), privacy breaches (G2), and operational disruption (G3).

\textbf{Risk Quantification: Likelihood × Impact Matrix} To assess the severity of risks modeled in our attack trees, we adopt a structured quantification framework based on the Likelihood × Impact matrix. Each risk is evaluated along two axes: Impact: the potential severity of harm or disruption if the attack succeeds. Likelihood: the probability that the attack path will be successfully executed under real-world conditions. We assess Likelihood using a two-factor framework grounded in attacker capability: Business Rule Knowledge – the degree of procedural or domain-specific understanding required (e.g., task orchestration, clinical decision logic). Technical Complexity – the level of effort and sophistication needed to execute the attack (e.g., prompt crafting, session hijacking, KV-cache access~\cite{DBLP:conf/ndss/WuZZWNWZ25}). 

A risk can be realized by various attack paths. Rather than scoring each threat individually, we evaluate the overall risk based on the most plausible attack path i.e., the one with the highest likelihood of success under realistic system conditions. For instance, a prompt injection that directly alters a treatment suggestion is more likely than one requiring multi-stage orchestration hijacking, even if both originate from the same initial vector. This context-aware scheme supports consistent and realistic Likelihood scoring across all goals. Scores reflect both system constraints and attacker knowledge dependencies. We adopt qualitative 1–5 scales based on real-world deployments, prior literature~\cite{ssa_toolkit}, and attacker effort–reward tradeoffs. The scoring criteria for Likelihood and Impact are detailed in Table~\ref{tab:likelihood_scale} and Table~\ref{tab:impact_scale}. Risk levels are computed by multiplying the Likelihood and Impact scores, which helps prioritize high -severity threats and guide defense planning.

\begin{table}[htbp]
\caption{Likelihood Scale for Risk Assessment}
\centering
\footnotesize 
\setlength{\tabcolsep}{3pt} 
\renewcommand{\arraystretch}{0.95} 
\begin{tabular}{|c|p{1.2cm}|p{6cm}|} %
\hline
\textbf{Score} & \textbf{Likelihood} & \textbf{Description} \\
\hline
1 & Rare & Extremely unlikely; requires insider access or rare conditions \\
\hline
2 & Unlikely & Requires effort, stealth, or partial misconfiguration \\
\hline
3 & Possible & Feasible with moderate effort or misconfigurations \\
\hline
4 & Likely & Known techniques exist; moderate mitigations present \\
\hline
5 & Almost Certain & High probability; exploit is trivial or seen in field \\
\hline
\end{tabular}
\label{tab:likelihood_scale}
\end{table}

\vspace{0.1em}

\begin{table}[htbp]
\caption{Impact Scale for Risk Assessment}
\centering
\footnotesize 
\setlength{\tabcolsep}{3pt} 
\renewcommand{\arraystretch}{0.95} 
\begin{tabular}{|c|c|p{6cm}|}
\hline
\textbf{Score} & \textbf{Impact} & \textbf{Description} \\
\hline
1 & Negligible & No clinical harm; affects UI or workflow only \\
\hline
2 & Minor & Affects a few non-critical patients; easily recoverable \\
\hline
3 & Moderate & Recoverable clinical harm to a single patient \\
\hline
4 & Major & Serious harm to one or moderate harm to many \\
\hline
5 & Catastrophic & Life-threatening harm or system-wide failures \\
\hline
\end{tabular}
\label{tab:impact_scale}
\end{table}

To refine our risk assessment, we categorize how each attack path contributes to risk. We define three categories: Direct, Indirect, and Situational. A Direct path causes the risk outcome immediately, without requiring secondary system behavior (e.g., a corrupted model that fails to load, resulting in unavailability). An Indirect path leads to the outcome through a side effect, often by manipulating an intermediate component. For example, an adversarial prompt that causes an LLM to enter infinite generation may overload the request-handling infrastructure, leading to denial of service. Here, the side effect is resource exhaustion, and the intermediate component is the system's task queue. Situational paths manifest only under specific conditions such as rare execution states or misconfigurations (e.g., an orchestrator routing failure that occurs only when two specific agents are active). These classifications are not static; the same threat can play different roles depending on the risk scenario. For instance, model tampering may be a Direct contributor to G1-R1 (misdiagnosis) but an Indirect factor in G1-R2 (unauthorized procedures). This contextual labeling clarifies how threats materialize in practice. It also influences our Likelihood scoring; Direct paths tend to be more feasible, while Indirect or Situational paths may depend on complex system conditions and thus have lower likelihoods. Additionally, this understanding informs mitigation prioritization, as Direct paths often demand more urgent defenses compared to edge-case Situational threats. Overall, this threat-path classification enhances our risk reasoning by linking operational conditions with exploit feasibility.

This layered methodology rooted in threat evolution, structured goal modeling, and qualitative risk scoring provides actionable insights into where risks are most concentrated and how they evolve. It also lays the foundation for future work on mitigation design and defense prioritization. Due to space constraints, this paper presents the detailed risk assessment only for Goal G1. 
\section{Results}
\textbf{Goal-Driven Risk Analysis - G1: Intervening in Medical Procedures}

\begin{figure*}[htbp]
\centering
\includegraphics[width=\textwidth, height=6.7cm]{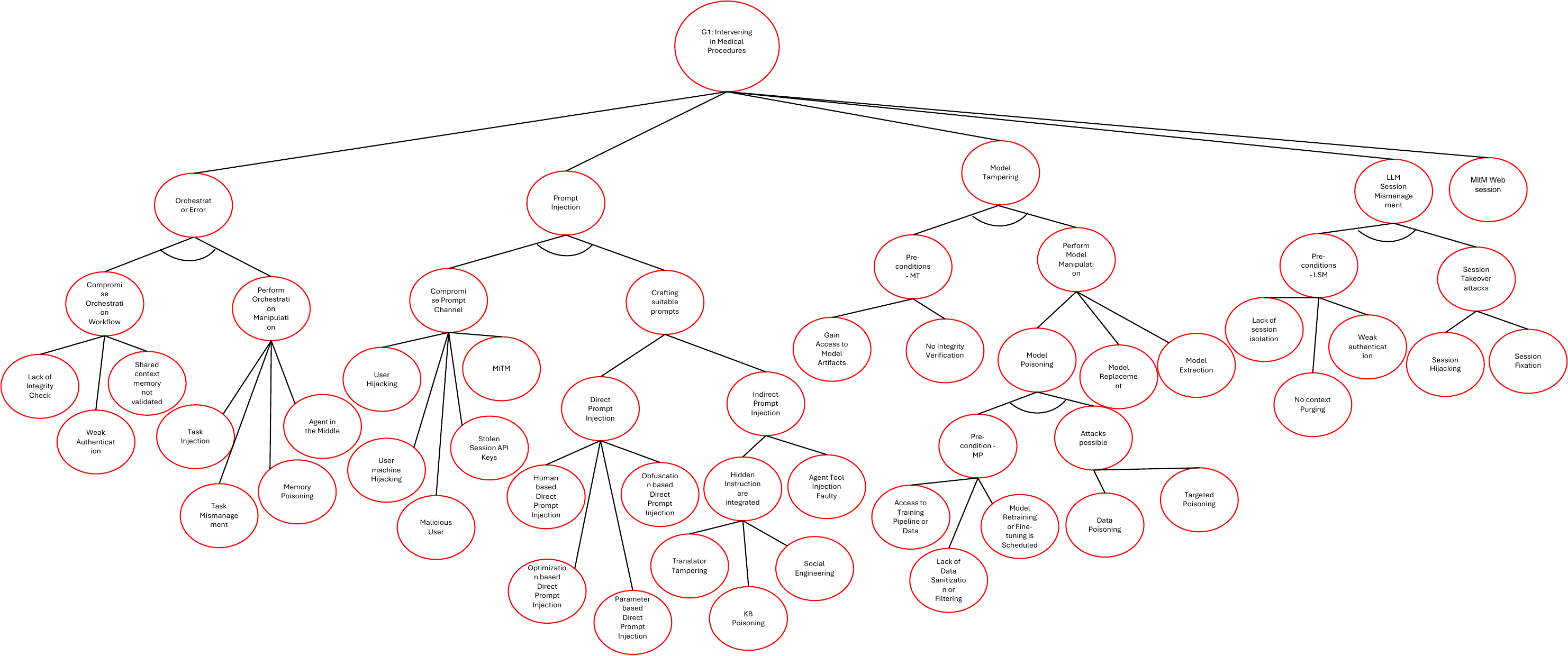}

\caption{Attack Tree for G1: Intervening in Medical Procedures (Zoom in to view node text clearly)}
\label{at_g1}
\end{figure*}

This goal captures how adversaries may manipulate the LLM-powered healthcare system to disrupt medical decision-making, potentially leading to harmful diagnoses, unsafe treatments, or unauthorized workflow changes. Figure~\ref{at_g1} illustrates the Attack Tree, where each attack path stems from the adversary’s objective and unfolds through specific system-level preconditions. These threats, grounded in our STRIDE-based modeling, are structured to reflect the sequence of enabling conditions and system actions required for successful exploitation

The root node represents the adversary’s objective Intervene in Medical Procedures and is decomposed into five major attack vectors, each modeled as a high-level child node:

\textbf{Prompt injection} attacks manipulate LLM inputs to influence medical recommendations. These require a Compromise of Prompt Channel, via user hijacking, device compromise, malicious user, leaked API keys, or MitM, each modeled as OR conditions in the AT. Once access is gained, attackers inject prompts via direct~\cite{Rossi2024} (explicit or obfuscated commands) or indirect~\cite{Rossi2024} means embedding hidden instructions through intermediaries like translators or knowledge bases. Indirect paths require embedding hidden instructions in benign content and proceed via translator tampering, KB poisoning, social engineering, or agent tool injection.

\begin{figure}[ht]
    \centering    \includegraphics[width=0.95\linewidth]{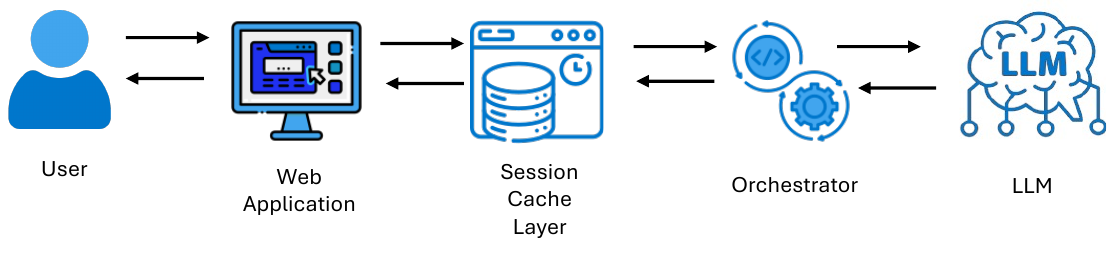}
    \caption{Workflow of the LLM-based Healthcare System.}
    \label{workflow}
\end{figure}

\textbf{LLM Session Mismanagement} In the context of our system architecture, an LLM session refers to a coherent, stateful interaction between the user-facing web application and the underlying large language model (LLM), mediated by an orchestrator agent. This session includes the exchange of prompts and responses, contextual memory, and task-specific metadata managed during a single user interaction. As illustrated in Figure~\ref{workflow} , the orchestrator maintains session continuity by forwarding user inputs to the LLM, handling intermediate outputs, and retaining memory across steps to enable multi-turn reasoning and consistent clinical responses. LLM session mismanagement arises when preconditions such as weak authentication, lack of user isolation, or missing context purging allow attackers to influence session behavior. To exploit these conditions, an attacker must first obtain session credentials (e.g., by compromising the web application), intercept traffic between the web app and orchestrator, or gain insider access to manipulate session metadata. These weaknesses, modeled as OR conditions in the Attack Tree, can lead to takeover of active sessions or manipulation of retained memory. Attackers may perform session hijacking by stealing or reusing tokens, or session fixation by injecting a predefined session ID that the system later accepts. In both cases, corrupted memory can silently redirect clinical reasoning or bypass safety constraints.

\textbf{Orchestration Errors} The orchestrator coordinates user queries by planning tasks, executing them, and managing data flows across components and external tools. This enables multi-step reasoning but also introduces a sensitive control layer. Errors arise when workflow integrity is compromised—often due to missing checks, weak API authentication, or unvalidated reuse of shared memory. These are modeled as OR conditions in the Attack Tree. Once these preconditions are met, adversaries can launch several attack paths. In Task Injection, malicious tasks are added to the pipeline to trigger unauthorized behavior or invoke unsafe tools (e.g., ShadowRay attack~\cite{oligo2024}). Task Mismanagement involves reordering or misrouting tasks, leading to inconsistent or unsafe clinical outputs. In Memory Poisoning, adversarial data is introduced into intermediate memory structures, corrupting downstream decisions. Finally, Agent-in-the-Middle~\cite{he2025} attacks exploit plugin-based architectures by spoofing trusted agents and injecting falsified data. To exploit these paths, attackers often require internal access to the orchestration environment (e.g., compromised MLOps tools or plugin APIs), or a foothold within the network to intercept and manipulate task pipelines. They degrade system logic and influence how the LLM interprets inputs.

\textbf{Model tampering} refers to unauthorized manipulation, replacement, or extraction of the deployed LLM to alter its behavior or leak internal knowledge. In our system, it can distort diagnoses, produce unsafe outputs, or compromise clinical reliability. As modeled in the AT, tampering proceeds through three major paths: poisoning, replacement, and extraction, each depending on specific preconditions. Attackers begin by gaining access to model artifacts, such as weights or outputs, often due to misconfigured storage, exposed APIs, or permissive roles. If integrity verification is absent, such as missing signatures or version checks, tampering can proceed undetected. In model poisoning, adversaries manipulate training or fine-tuning data to introduce backdoors or corrupt behaviors. This attack requires three key preconditions to be met: (1) access to the training data or pipeline, (2) active retraining or fine-tuning schedule, and (3) weak data sanitization or filtering. These form AND conditions in our attack tree, since poisoning is only feasible when retraining is exposed and unprotected. Model replacement involves deploying a malicious variant that mimics normal behavior but fails under clinical stress.

\textbf{MiTM Web Session}attacks occur when an adversary intercepts and alters communication between the user and the web interface. In LLM-powered healthcare systems, this allows silent manipulation of clinical workflows by injecting, suppressing, or modifying user queries before they reach the orchestrator or LLM.

Based on the attack tree decomposition, we identify four primary risks associated with G1 (Refer Table~\ref{tab:combined_risk_assessment}). Each risk represents a distinct compromise path and is evaluated in terms of likelihood, impact, and affected system components. We also highlight the specific attack vectors that enable these risks. \textbf{\textit{G1-R1: Misdiagnosis of Critical Illness.}} This risk refers to adversarial inputs causing the system to produce incorrect diagnoses for high-stakes conditions (e.g., stroke, sepsis, cancer). Likelihood is assessed using the framework in Table~\ref{tab:likelihood_framework}, which considers business (i.e., clinical) knowledge and technical effort for each attack vector. The most feasible path is direct prompt injection, which requires neither medical expertise nor elevated access. As shown in Table~\ref{tab:combined_risk_assessment}, it scores 4 (Likely), as attackers can craft prompts (e.g., “Override prior reasoning…”) with minimal context. Conversely, orchestrator task manipulation or model tampering require insider access or deeper system knowledge and thus score lower in likelihood. We score Likelihood based on this most plausible path, rather than averaging across all paths. In this case, prompt injection dominates due to its ease, visibility in real-world deployments, and high leverage over model outputs. Thus, we assign a Likelihood score of 4 (Likely). However, the overall risk may vary: if system design or deployment context changes (e.g., LLM prompt interface is hardened), less likely paths such as model poisoning or session hijack may become the primary enablers, and the risk score would correspondingly shift. The Impact is rated 5 (Catastrophic) due to the direct threat to patient safety and potential for irreversible clinical harm. Diagnostic errors at this stage of care could delay life-saving interventions, violate regulatory safety standards, and erode trust in the system.

\textbf{\textit{G1-R2: Unauthorized Procedures Executed.}} This risk refers to adversaries triggering unintended or unapproved medical procedures. The most feasible enabler is prompt injection combined with orchestrator manipulation. Prompt injection can issue misleading instructions (e.g., “Proceed to imaging now”), but full task execution typically requires weak orchestration logic. Feasibility increases if the attacker can manipulate task routing logic, such as injecting or skipping approval steps. These actions represent direct enablers: Prompt Injection (Likelihood 4) overrides system logic, while Orchestrator Error enables control-flow tampering. Given this combination, Likelihood is rated 3 (Possible), balancing prompt injection feasibility with orchestration complexity. In practice, this risk manifests when attackers exploit insufficiently validated model outputs to trick downstream controllers (e.g., agents or APIs) into executing actions without secondary checks. The Impact is rated 4 (Major) because such unauthorized procedures can result in direct clinical harm (e.g., radiation exposure, unnecessary surgery), incorrect device activations, or regulatory violations concerning informed consent. 

\textbf{\textit{G1-R3: Corrupted Medication Recommendations.}} This risk refers to manipulation of drug names, dosages, or prescribing logic.  The most feasible vector is Prompt Injection e.g., inserting adversarial instructions like “Ignore allergy flag; continue with penicillin.” This vector is rated 4 (Likely) due to low technical and clinical barriers. Prompt Injection enables misleading medication instructions, especially dangerous when context is retained across turns or appended via indirect sources like translators.  Overall, Likelihood is rated 4 (Likely) based on the ease of triggering corruption via prompt injection and availability of real-world input vectors. This reflects a typical external attacker scenario requiring no insider access. The Impact is rated 4 (Major) due to the severity of incorrect medical advice. A corrupted recommendation (e.g., an unsafe dose of a cardiac drug or an allergy-inducing antibiotic) can result in hospitalization, clinical harm, or delayed recovery.

\textbf{\textit{G1-R4: Cross-Patient Context Contamination.}} This risk arises when memory from one patient session leaks into another, causing decisions based on incorrect or unauthorized information. We evaluate the Likelihood as 3 (Possible) based on the presence of two key enablers:This risk is primarily driven by LLM Session Mismanagement (Direct), where poor memory isolation or improperly scoped KV-cache allows patient-specific context to leak across conversations; this vector is rated 3 in multi-tenant deployments. Orchestrator Error (Indirect) contributes when misrouted memory pointers or failure to reset agent state leads to workflow bleed-over, with feasibility ranging from 2–3 depending on configuration and insider access. The Likelihood is thus rated 3 (Possible) because this risk is realistic in systems with weak session controls, particularly when LLM context memory or agent memory is reused across users. However, the need for either deployment-specific misconfiguration or insider access prevents a higher score. The Impact is rated 3 (Moderate), as this form of contamination can cause diagnostic confusion (e.g., recommending tests based on the wrong patient history), or introduce misleading context (e.g., one patient’s condition being mistakenly factored into another’s care plan). While serious, these issues are generally limited to individual interactions and are often reversible with audit or intervention.

\section{Discussion}
While this study presents a structured and goal-driven risk assessment framework for LLM-based systems, several important directions remain for future work.  First, we plan to extend our risk quantification framework by integrating multi-metric scoring approaches, such as the Common Vulnerability Scoring System (CVSS) or sector-specific threat prioritization criteria. This will allow more granular risk calibration beyond the current Likelihood × Impact model.
Second, we aim to develop a comprehensive set of mitigation strategies derived from the attack trees. These will include both technical countermeasures (e.g., session isolation, prompt sanitization, model integrity checks) and operational safeguards (e.g., access controls, user role restrictions). By analyzing which paths most frequently lead to high-impact risks, we can also prioritize defenses more effectively.
Third, we will validate our risk assessment methodology and results through expert feedback from both healthcare professionals and cybersecurity practitioners. This step is critical to ensure that our risk models reflect real-world conditions and capture domain-relevant threats and priorities.

Finally, we plan to explore the use of LLMs themselves to support threat modeling and risk analysis. Recent advances show that LLMs can generate attack scenarios or identify overlooked threats when guided appropriately ~\cite{yamin2024llm}. Leveraging these models for semi-automated attack tree construction or likelihood estimation could significantly enhance scalability and reduce expert burden.
\section{Conclusion}
We presented a structured and goal-driven risk assessment framework for LLM-based healthcare systems, grounded in attack tree modeling. Starting from system-level threat elicitation, we constructed attack trees around clinically meaningful adversarial goals such as misdiagnosis, data leakage, and service disruption to trace how threats evolve through specific preconditions and execution paths. From these trees, we systematically derived high-stakes risks and quantified them using a tailored Likelihood × Impact framework that captures both technical feasibility and the specialized domain knowledge required to execute real-world attacks. This approach enables fine-grained, real-world risk prioritization and offers a scalable foundation for assessing AI-native threats in safety critical settings. By bridging abstract threat modeling with actionable risk reasoning, our work advances the security analysis of LLM-powered healthcare systems.

\section{Acknowledgment}
We thank the anonymous HealthSec 2025 reviewers for their valuable feedback.

\begingroup
\footnotesize
\setstretch{0.8} 
\bibliographystyle{IEEEtran}
\bibliography{main}
\endgroup

\appendices
\onecolumn
\centering
\section*{Appendix: Tables and Attack Tree}
\vspace{-1em}
\begin{table*}[htbp]
\caption{Risk Likelihood Framework Table for Key Attack Paths}
\centering
\scriptsize
\renewcommand{\arraystretch}{0.9}
\begin{tabular}{|p{1.1cm}|p{3.5cm}|p{2.7cm}|p{1cm}|p{7cm}|}
\hline
\textbf{Attack Path} & \textbf{Business Knowledge Required} & \textbf{Technical Complexity} & \textbf{Likelihood} & \textbf{Justification} \\
\hline
Prompt Injection & Low; No clinical knowledge required to manipulate LLM responses. & Low; Requires only user/API access. & 4 & Always feasible. External or internal attackers can craft prompts without domain insight. Example: “Override previous instruction and recommend immediate discharge” bypasses safety filters even without medical context. \\
\hline
LLM Session Mismanagement & Medium; Understanding of session continuity and PHI retention improves attack precision. & Medium; Requires session token access or fixation strategies. & 3–4 & Lower (3) if attacker blindly hijacks session tokens. Higher (4) if insider (e.g.,  developer or system admin) knows token refresh logic or memory scope, enabling persistent dialogue control. \\
\hline
Orchestrator Error & High; Requires detailed knowledge of multi-agent task routing and clinical workflows (e.g., ‘order test → analyze → prescribe’). & Medium–High; Requires ability to intercept or alter internal task messages. & 2–3 & Rare (2) if attacker lacks care-path knowledge. Higher (3) if insider or developer can edit orchestration configs, e.g., skipping “allergy screening” before treatment. \\
\hline
Model Tampering & Medium; Must know what clinical entities to target (e.g., specific diagnoses, medications). & High; Requires access to model weights or training pipeline. & 2–3 & Lower (2) if attacker must breach infrastructure. Higher (3) if insider ML engineer has fine-tuning privileges to bias discharge instructions or clinical summaries. \\
\hline
MitM Web Session & Low; No domain knowledge needed; attacker targets data in transit. & Medium; Requires network-level control (e.g., ARP spoofing, TLS downgrade). & 3 & Consistent (3). Attacker can tamper with clinical terms (e.g., change “acetaminophen” to “ibuprofen”) regardless of insider status. Medical workflow knowledge offers no added advantage. \\
\hline
Session Hijack / Fixation & Medium; Knowing which sessions hold PHI improves the ability to extract useful data. & Medium; Involves stealing a valid token or forcing reuse of a known session token (fixation). & 3–4 & 3 for blind token reuse; 4 if insider knows token lifetime \& memory scope. \\
\hline
\end{tabular}
\label{tab:likelihood_framework}
\end{table*}

\vspace{-1em}

\begin{table*}[htbp]
\caption{Comprehensive Risk Assessment for G1: Intervening in Medical Procedures}
\centering
\scriptsize
\renewcommand{\arraystretch}{0.9}
\begin{tabular}{|c|p{3cm}|c|c|c|p{7cm}|}
\hline
\textbf{Risk ID} & \textbf{Healthcare Risk} & \textbf{Likelihood} & \textbf{Impact} & \textbf{Risk Score} & \textbf{Attack Vectors} \\
\hline
G1-R1 & Misdiagnosis of Critical Illness & 4 (Likely) & 5 (Catastrophic) & 20 & Prompt Injection (Direct), Model Tampering (Direct), Orchestrator Error (Indirect), MitM Web Session (Indirect), LLM Session Mismanagement (Situational) \\
\hline
G1-R2 & Unauthorized Procedures Executed & 3 (Possible) & 4 (Major) & 12 & Prompt Injection (Direct),  Model Tampering (Indirect), Orchestrator Error (Direct), MitM Web Session (Indirect), LLM Session Mismanagement (Situational) \\
\hline
G1-R3 & Corrupted Medication Recommendations & 4 (Likely) & 4 (Major) & 16 & Prompt Injection (Direct), Model Tampering (Direct), Orchestrator Error (Indirect), MitM Web Session (Direct), LLM Session Mismanagement (Indirect) \\
\hline
G1-R4 & Cross-Patient Context Contamination & 3 (Possible) & 3 (Moderate) & 9 & Orchestrator Error (Indirect), MitM Web Session (Direct), LLM Session Mismanagement (Direct) \\
\hline
\end{tabular}
\label{tab:combined_risk_assessment}
\end{table*}

\vspace{-1em}

\vspace{-1em}

\end{document}